\documentstyle[aps,prb,psfig]{revtex}
\begin{document}
\twocolumn[\hsize\textwidth\columnwidth\hsize\csname @twocolumnfalse\endcsname
\title{Electron-hole correlation effects in the emission of light from quantum
wires}

\author{Francesco Tassone}
\address{Ginzton Laboratory, Stanford University, Stanford, CA-94305,
and ERATO Quantum Fluctuation Project} 

\author{Carlo Piermarocchi} \address{Institut
de Physique Th\'{e}orique, Ecole Polytechnique F\'{e}d\'{e}rale,
CH-1015 Lausanne, Switzerland}
 \date{\today}

\maketitle

\begin{abstract}
We present a self-consistent treatment of the electron-hole
correlations in optically excited quantum wires within the ladder
approximation, and using a contact potential interaction. The
limitations of the ladder approximation to the excitonic low-density
region are largely overcome by the introduction of higher order
correlations through self consistency. We show relevance of these
correlations in the low-temperature emission, even for high
density relevant in lasing, when large gain replaces excitonic
absorption.
\end{abstract}


] 

Based on the singularity of the density of states (DOS) of massive
particles at the band edge in one dimension (1D) and the single
particle picture, large gains and significant improvements in optical
devices have been predicted for 1D systems.\cite{arakawa,dos2}
However, inclusion of the Coulomb correlations among the photoexcited
carriers results in a {\em finite} absorption at the band gap, i.e. in
a vanishing Sommerfeld factor.\cite{ogawa,rossi} This effect is
related to the large exciton binding energy and oscillator strength,
which strongly reduces that available for free carriers. Correlation
between electrons and holes is therefore very important in 1D, and
likely to persist up to large carrier densities. It has been shown
that screening in 1D systems has a little influence on optical
nonlinearities,\cite{benner91} and in a first approximation can be
neglected. Thus, a minimal model describing the optical properties of
the wires has to include the strong Coulomb correlation between
electron and holes, both in the low density excitonic limit and for
the higher densities, when the Fermi gas becomes
degenerate. Therefore, the time dependent Hartree-Fock (HF)
approximation for the electron-hole system, which is used in the
derivation of the semiconductor Bloch equations, is not sufficient for
this purpose.  Indeed, this fact is also recognized for higher
dimensional systems, where, beside screening effects, the
time-dependent HF was improved by including scattering between
carriers at the Born level.\cite{lindberg} But, even if in the optical
response the exciton resonance is reproduced, the excited electrons
and holes in this model are still the single particle states of the
Hartree-Fock theory, thus bound electron-hole pairs are not included.
In this work we go beyond these approximations, and include
correlation effects by summing up electron-hole scattering to infinite
order in ladder-type diagrams (ladder approximation, LA). The
resulting renormalized interaction (T-matrix) is known to correctly
reproduce the bound states at low carrier densities, and Coulomb
correlations among the unbound states. Phase-space filling effects are
also partially included.\cite{zimmi} However, the range of validity of
the LA is in fact rather limited, as higher order correlations become
important at higher densities. These can be partially recovered within
a self-consistent approach (self consistent ladder approximation,
SCLA), where the bare electron and hole propagators are dressed to
include the effects of scattering with other carriers on
propagation. In this way, multiple scattering between bound and
unbound pairs are effectively included.\cite{haussman} These
correlations are relevant for temperatures smaller than the exciton
binding energy, which is significantly large in these systems. In this
work we consider a simplified coulomb interaction potential in order
to clearly assess this statement and compare its results to simpler
models, which do not include correlation.  As an interesting example,
we finally show that emission calculated within our model
shows qualitative agreement with experimental observations,\cite{devaud}
explaining the origin of apparent absence of band gap renormalization in the
emission for the wires.

We here consider one parabolic, and spinless, conduction band for
electrons (creation operators $c_k$) and one for holes ($d_k$), with
the same masses, interacting through a simple contact potential
$V(x)=a\delta(x)$, where $a=2 E_b a_B$, $E_b$ and $a_B$ being the
binding energy and the Bohr radius of the exciton. Electrons and holes
are independent fermions, and the corresponding operators commute. We
introduce dimensionless units, where $E_b=1/2$, and lengths are scaled
with the 1D Bohr radius. Then, the Hamiltonian reads $${\cal H}=\sum_k
\frac{k^2}{4}\left[c_k^\dagger c_k+d_k^\dagger d_k\right]-
\sum_{k,k',q} c_{k-q}^\dagger d_{k'+q}^\dagger d_{k'} c_k.$$ The
choice of the contact potential makes T-matrix self-consistency
manageable to both numerical and analytical treatment. Within this
potential, electron-electron and hole-hole interactions are absent for
the same spin. Although a drawback for realistic calculations, it
allows us to focus on the electron-hole correlation alone. We also
remark that screening can not be consistently introduced within such a
restricted form of the potential. But this is presumably a minor
drawback in 1D, as remarked above. Standard finite temperature Green
function techniques may be used to write the SCLA. However, the
solution of the self-consistent problem is by necessity numerical, and
the analytical extension to real frequencies is numerically difficult,
due to the complex structure of the resulting Green functions
.\cite{scalapino} For this reason we work directly with real
frequencies/times, where the Green functions and self-energies are
matrices defined on the Keldysh contour,\cite{danielewicz} and their
matrix elements related by the Kubo-Martin-Schwinger relations for
thermal equilibrium.  The full SCLA reads: $$ T_k(\omega)=1+\int_C
d\omega'\sum_{q'} \left[i G_{k-q'}(\omega')
G_{q'}(\omega-\omega')\right] T_k(\omega)$$ \begin{equation}
G_{k}(\omega)=G^{(0)}_{k}(\omega)+G^{(0)}_{k}(\omega)
\Sigma_{k}(\omega) G_{k}(\omega)\label{scla}\end{equation} $$
\Sigma_{k}(\tau)=\sum_{q} i T_k(\tau)G_{q-k}(-\tau).\label{scla3} $$
The SCLA is a {\em conserving approximation}\cite{KB}: particle number
and energy are conserved, making the SCLA robust at any density, even
when large scattering produce significant renormalizations. A detailed
study of SCLA for both the attractive and repulsive (single-band)
Hubbard model at half filling shows that the SCLA is substantially
close to the available exact solution in an extensive range of
temperatures/magnetizations.\cite{buzatu} We expect this to be even
more true for this case of negligible fillings (continuum model).

In the low density limit, using the bare particle propagators, the
$T$-matrix shows a pole at the exciton energy,
$T^+(k,\omega)\simeq[\omega-(-1/2+k^2/8)+2i\gamma]^{-1}$. At the next
self-consistency step, the particle propagators show structures below
the exciton energy through the self-energy $\Sigma$. Higher order
terms in self-consistency then introduce exciton-exciton scattering
effects.\cite{haussman} These effects are actually overestimated for the
contact potential with respect to a realistic coulomb potential due to
the lack of electron-electron (and hole-hole) interactions in the
spinless model.\cite{note1} We also checked that inclusion of spin in
the SCLA (using the contact potential)  reduces scattering 
for a given density.  In order to visualize the effects of
self-consistency on the single particle Green function, we plot in
Fig. 1 the spectral function at different steps of iteration for
T=0.1. At step 0, the spectral function shows the usual single
particle single peak, broadened by $\gamma$=0.02 for
regularization. At step 1, excitonic correlations appear. Due to
finite temperature, not only a single excitonic peak appears, but
contributions from larger exciton momenta are also evident. The
discretization of the k-space (16 points) also results in a
fragmentation of this two particle continuum into resolved peaks, and
in a reduction of the binding energy at low densities from 0.5 to
0.37. Further self-consistency steps significantly broaden and {\em
shift} both the single particle and two-particle peaks. We thus
understand that interaction among carriers is shifting both exciton
energies and single particle energies.

\begin{figure} 
\centerline{\psfig{figure=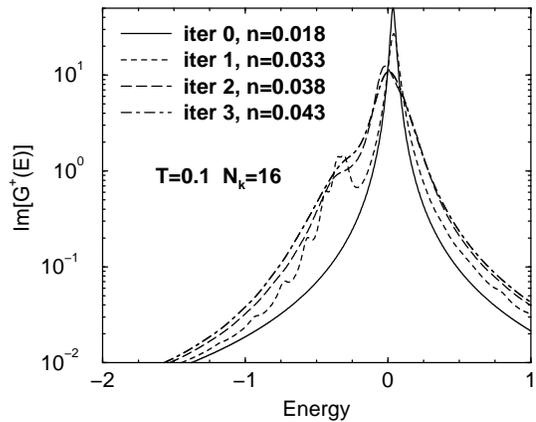,width=7truecm}}
\caption{The spectral function $\Im[G^+]$ at k=0, first four steps of
iteration. Corresponding densities are also shown. Other
parameters in the figure.}
\end{figure}

Next, we introduce the optically measurable quantities, emission (or
photoluminescence), and absorption. We calculate them in the dipole
approximation, and neglecting retardation in the electromagnetic
 interaction between carriers (polaritonic effects). The polarization
$P$ is the linear response to the classical external transverse field,
calculated within the dipole approximation. It is thus related to the
two-particle electron-hole Green function, where the external electron
and hole lines are closed over the interaction with the transverse
photon:
$$ P({\bf k})\propto \sum_{ \bf q,Q} 
G_2(\frac{\bf k}{2}-{\bf q},\frac{\bf k}{2}+{\bf q};\frac{\bf k}{2}-{\bf q}-{\bf Q},\frac{\bf k}{2}+{\bf q}+{\bf Q}) = $$
\begin{equation}
\label{pol}H({\bf k})+H({\bf k})T({\bf k})H({\bf k}),\end{equation}
with $H({\bf k})=i\sum_{\bf q} G({\bf k}/2-{\bf q})G({\bf k}/2+{\bf
q}).$ Here we used the notation ${\bf k} = (k,\omega)$, and optical
frequencies are measured from the gap.  The first term describes the
usual band to band recombination found in the single particle picture,
whereas the other term includes the correlation effects of bound
electron-hole states.  Here we notice that Eq. (\ref{pol}) gives the
linear response of the system to the external field only when the
exact $G_2$ is used.\cite{KB} However, we may expect that it is
accurate also in the SCLA as we noticed above that the SCLA is
expected to be quantitatively close to the exact solution. We checked
that deviations from current-conservation in the linear response are
negligibly small. 

 The absorption $\alpha$, related to ${\Im} [P^+]$, (the $+$ apex
labelling the retarded function), can be written as
$$ \alpha({\bf k})\propto {\Im} [H^{+}({\bf k})] |T^{+}({\bf k})|^2.
$$
Thus, absorption in the
excitonic limit shows a pole at the exciton energy at low
densities. Moreover, we may also analytically find the absorption
above the gap by using the analytic expression of
$H^{+}(k=0,\omega)=-i/\sqrt{2\omega +i0^+}$ for small densities: 
$$\alpha(k=0,\omega)=\frac{\sqrt{2\omega}}{1+{2\omega}}$$ which clearly shows
cancellation of the 1D DOS divergence at $\omega\rightarrow 0^{+}$, i.e. a
Sommerfeld factor which vanishes linearly at $\omega=0$.
The $k=0$ corresponds to the direction perpendicular to the wire.
Emission is instead related to the correlation function of the dipole operator
and therefore to $P^{<}(k,\omega)$. It can be found from absorption
through the Kubo Martin Schwinger relations:
$$P^{<}(k,\omega)\propto
\left[\exp{\beta(\omega-2\mu)}-1\right]^{-1} \Im [P^{+}(k,\omega)]
$$
which show the typical {\em Bose} occupation factor for the
emission. We conclude that at low densities/temperatures, emission
is excitonic as described in absorption by the poles of $T^{+}$.

\begin{figure} 
\centerline{\psfig{figure=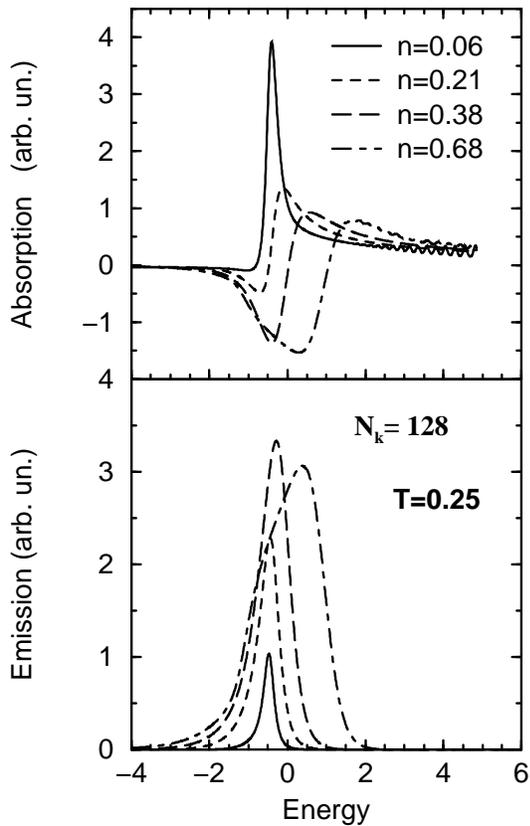,width=7truecm}}
\caption{The absorption and emission  at various
densities. Other parameters in the figure.}
\end{figure}

In the following, we discuss emission and absorption at intermediate
and high densities, when a relevant number of unbound pairs coexist
with bound pairs. This situation is relevant for the photoluminescence
experiments, and for low temperature laser operation, when hot
carriers are injected in the system. Of course, quite stronger
correlation effects can be found at smaller temperatures, when most of
the carriers are bound into excitons, but our aim is to show the
relevance of correlation in a typical and less extreme situation. We
solved Eq. (\ref{scla}) for T=0.25 (corresponding to $E_b/2$) and
densities up to n=0.8.  In Fig. 2 we plot the resulting absorption and
emission. The low density limit shows the excitonic absorption, and
the band to band absorption, as a long tail at higher energy. At
larger densities, excitonic absorption is bleached and gain becomes
evident. Even for n=0.38, when large gain is present in the exciton
spectral region, the emission peak stays fixed close to $E_b\sim
-1/2$, the low-density exciton energy. At the largest densities
n$>$0.4, the emission intensity saturates and a flattened top of the
emission peak results from the significant depth of the Fermi sea of
electrons and holes.  We notice that even for n$<0.4$, significant
broadenings are predicted by the SCLA. These are instead absent in the
HF calculations and have to be introduced either phenomenologically,
or by treating scattering among carriers at the Born level. We remark
that scattering at the Born level is just the first of the infinite
terms in the ladder expansion.  In order to better understand the
origin of the stability of the energy of the emission peak with
density, we compared the exciton energy, the emission energy, and the
band gap renormalization (BGR), defined as twice the shift of the
single particle energy (Fig. 3). This definition coincides with the
optical band gap renormalization at small densities.  We also defined
the exciton energy as that of the pole of the T-matrix (more
precisely, the $|T^+|^2$ peak).  
\begin{figure}
\centerline{\psfig{figure=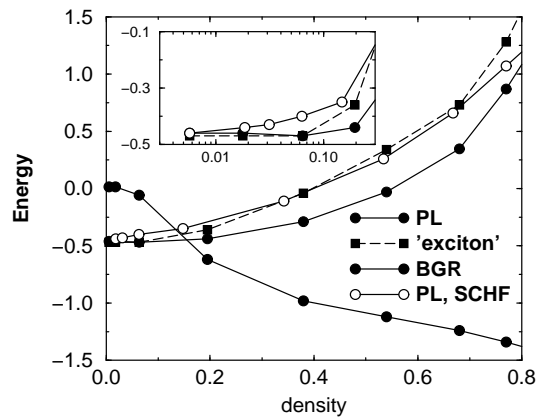,width=7truecm}}
\caption{Emission peak energy (PL), emission peak energy in HF (PL-SCHF), 
the exciton energy ($|T^+|^2$ peak), and band gap renormalization
(BGR) as functions of the density. In the inset an enlargement for the
lower densities.}
\end{figure}
In Fig. 3 we notice large {\em
negative} BGR even at relatively small density. For $n\sim 0.15$, the
BGR is comparable to $E_b$, but a small change in the exciton energy
is observed (Fig. 3, inset). We remark that the values of the BGR are
comparable to those calculated using a more realistic coulomb
potential within the random phase approximation.\cite{dassarma} It
confirms that screening effects on BGR are small for low densities,
when the depth of the Fermi sea is smaller than the temperature as in
our case. Intuitively, in our model BGR and binding energy compensate
to produce a negligible shift of the exciton energy. However, this
compensation goes well beyond the densities where the concept of
exciton as a bound state is still relevant, i.e. when the bandgap
falls below the exciton energy, at $n>0.15$. Here, the T-matrix still
shows a peak. It is now originating from a two-particle scattering
resonance in the two-particle continuum.  From Fig. 3, we notice an
agreement between the exciton energy and emission (PL) peak for
$n<0.2$. It' s remarkable that the emission peak does not shift much
even for $0.2<n<0.4$, when the exciton energy does. But, this
discrepancy between the two appears in a region where significant
broadening of both the peaks is present, as clearly shown in Fig. 2
for the emission.  This stability of the emission peak is not
correctly predicted by the simpler self-consistent HF approximation
(using the same contact potential for consistency), also shown in
Fig. 3. Even at lower densities (Fig. 3, inset), when self-consistent
HF correctly predicts excitonic emission, larger emission shifts are
found compared to the SCLA. Moreover, we again remind that the HF
approximation does not consistently introduce broadenings in the
spectra. Only at densities $n>0.8$ the self-consistent HF and the SCLA
coincide, given the larger weight of uncorrelated carriers in the
system at these large densities. We here remark that the same fixed
emission peak for a broad range of densities was also observed in
recent experiments on high quality GaAs quantum wires at comparable
temperatures, and attributed to excitonic emission.\cite{devaud} From
this emission data, the authors also concluded that BGR is
negligible. Here, we show instead that emission data is not sufficient
to support this conclusion, and that indeed BGR can coexist with a
fixed emission peak in the 1D system. We also remind that our
simplified model is {\em overestimating} broadenings and shifts, so
that this conclusion holds even more true.

In conclusion, we have shown the importance of electron-hole
correlation effects in the emission of light from quantum wires, even
at large densities, when the temperature is smaller than the exciton
binding energy. We discussed how the self-consistent ladder
approximation includes a large part of these correlations, and in
particular the broadening and renormalization resulting from multiple
scattering between correlated pairs. It is able to predict negligible
shifts of the emission peak in a large range of densities, in
agreement with experimental observations, and in contrast to the
predictions of simpler Hartree-Fock approximation. Moreover, we
observed that this feature coexists with significant band-gap
renormalization, and even with the disappearance of the excitonic peak
in absorption. The SCLA is thus the simplest model retaining
sufficient correlations to explain gain and emission characteristics
of the wires at low temperatures, even in the high-density 
regime relevant for lasing.

We thank R. Ambigapathy, 
C. Ciuti, 
B. Deveaud, 
A. Quattropani,
V. Savona, 
P. Schwendimann, 
P. E. Selbmann, 
Y. Yamamoto, and 
R. Zimmermann, for many stimulating
discussions. The present work has been supported by the
Swiss National Science Fundation, by the Swiss National Priority
Program for Optics, and by the ERATO Quantum Fluctuation Project.


\begin{references}
\bibitem{arakawa} Y. Arakawa and H. Sakaki, Appl. Phys. Lett. {\bf 40}, 939, (1982).

\bibitem{dos2} E. Kapon, Proc. IEEE {\bf 80}, 398, (1992).

\bibitem{ogawa} T. Ogawa and T. Takagahara, Phys. Rev. B {\bf 43}, 14325, 
(1991).

\bibitem{rossi} F. Rossi and E. Molinari,  Phys. Rev. Lett. {\bf 79}, 3642, (1996).

\bibitem{benner91} S. Benner, and H. Haug, Europhys. Lett. {\bf 16}, 579 (1991).

\bibitem{lindberg} M. Lindberg and S.W. Koch, Phys. Rev. B {\bf 38}, 3342,
 (1988).

\bibitem{zimmi} for a review see e.g., R. Zimmermann, Many Particle Theory of Highly Excited Semiconductors, (BSB Teubner, Liepzig, 1987).

\bibitem{haussman} R. Haussman, Z. Phys. B {\bf 91}, 291, (1993).

\bibitem{devaud} R. Ambigapathy {\it et al.}, Phys. Rev. Lett. {\bf 78}, 3579, 
(1997).

\bibitem{scalapino} S. R. White {\it et al.}, Phys. Rev. Lett. {\bf 63}, 1523, 
(1989).

\bibitem{danielewicz} P. Danielewicz, Annals of Physics {\bf 152}, 239, 
(1984).

\bibitem{KB} G. Baym and L. P. Kadanoff, Phys. Rev. {\bf 124}, 287, (1961).

\bibitem{buzatu} F. Buzatu, Modern Physics Letters B {\bf 9}, 1149, (1995).

\bibitem{note1} F. Tassone and C. Piermarocchi, unpublished.

\bibitem{dassarma} Ben Yu-Kang, and S. Das Sarma,  Phys. Rev. B {\bf 48}, 5469, (1993).

\end{references}
\end{document}